\documentclass[12pt]{article}
\usepackage{axodraw,epsfig}
\usepackage{graphics}

\catcode`@=11
\newcount\@tempcntc
\def\@citex[#1]#2{\if@filesw\immediate\write\@auxout{\string\citation{#2}}\fi
  \@tempcnta\z@\@tempcntb\m@ne\def\@citea{}\@cite{\@for\@citeb:=#2\do
    {\@ifundefined
       {b@\@citeb}{\@citeo\@tempcntb\m@ne\@citea\def\@citea{,}{\bf ?}\@warning
       {Citation `\@citeb' on page \thepage \space undefined}}%
    {\setbox\z@\hbox{\global\@tempcntc0\csname b@\@citeb\endcsname\relax}%
     \ifnum\@tempcntc=\z@ \@citeo\@tempcntb\m@ne
       \@citea\def\@citea{,}\hbox{\csname b@\@citeb\endcsname}%
     \else
      \advance\@tempcntb\@ne
      \ifnum\@tempcntb=\@tempcntc
      \else\advance\@tempcntb\m@ne\@citeo
      \@tempcnta\@tempcntc\@tempcntb\@tempcntc\fi\fi}}\@citeo}{#1}}
\def\@citeo{\ifnum\@tempcnta>\@tempcntb\else\@citea\def\@citea{,}%
  \ifnum\@tempcnta=\@tempcntb\the\@tempcnta\else
   {\advance\@tempcnta\@ne\ifnum\@tempcnta=\@tempcntb \else \def\@citea{--}\fi
    \advance\@tempcnta\m@ne\the\@tempcnta\@citea\the\@tempcntb}\fi\fi}
\catcode`@=12

\begin{document}

\begin{titlepage}

    \begin{flushright}
      \normalsize PITHA~08/07\\
      \normalsize TTP/08-11\\
      \normalsize SFB/CPP-08-18\\
      April 22, 2008 
    \end{flushright}

\vskip1.5cm
\begin{center}
\Large\bf\boldmath
 Ultrasoft contribution to heavy-quark pair production near threshold
\unboldmath
\end{center}

\vspace*{0.8cm}
\begin{center}

{\sc M. Beneke}$^{a}$ and 
{\sc Y. Kiyo}$^{b}$\\[5mm]
  {\small\it $^a$ Institut f{\"u}r Theoretische Physik~E,\\
    RWTH Aachen University,}\\
  {\small\it D--52056 Aachen, Germany}\\[0.2cm]
  {\small\it $^b$ Institut f{\"u}r Theoretische Teilchenphysik,
    Universit{\"a}t Karlsruhe,}\\
  {\small\it D--76128 Karlsruhe, Germany}
        
\end{center}

\vspace*{0.8cm}
\begin{abstract}
  \noindent
  We compute the third-order correction to the heavy-quark 
  current correlation function due to the emission and
  absorption of an ultrasoft gluon. Our result supplies a missing 
  contribution to top-quark pair production near threshold 
  and the determination of the bottom quark mass from 
  QCD sum rules.

\vspace*{0.8cm}

\end{abstract}

\vfil
\end{titlepage}

\newpage


In a previous paper \cite{Beneke:2007pj} we described the calculation
of the ultrasoft gluon contribution to the correlation function of 
heavy-quark vector currents, which constitutes one of the major
missing pieces in perturbative calculations of quarkonium-like 
systems at the third order in non-relativistic perturbation theory. 
That paper presented the result for the residues of the bound-state 
poles of the correlation function, which relate to quarkonium decay 
constants. (The ultrasoft contribution to the $S$-wave quarkonium
masses  has been known for some time \cite{KniPen1}.) However, 
some of the most interesting physics related to 
the heavy-quark threshold -- the determination of the bottom quark 
mass from sum rules \cite{NSVZ} and the top-quark pair-production 
cross section \cite{FK87} -- requires the calculation of the full 
energy-dependent correlation function (see also the 
reviews \cite{Beneke:1999zr,Hoang:2000yr}). In this Letter we 
provide the result for the ultrasoft correction to the full 
correlation function.

The effective field theory formalism and technical set-up for 
this calculation have already been given in \cite{Beneke:2007pj} 
and will not be repeated in detail here. In brief, we consider 
the current correlation function 
\begin{equation}
2 (d-1) N_c \,G(E) = 
i\int d^dx\,e^{iE x^0}\,\langle 0|T [\chi^\dagger\sigma^i\psi](x)
\,[\psi^\dagger\sigma^i\chi](0)|0\rangle \,,
\label{defG}
\end{equation}
where $E=\sqrt{s}-2m$ is small, $\sqrt{s}$ is the centre-of-mass energy and 
$m$ the heavy-quark pole mass. ($d=4-2\epsilon$ is the space-time 
dimension in dimensional regularization.) The heavy-quark current 
$\psi^\dagger\sigma^i\chi$ is defined in non-relativistic QCD. 
The (leading) ultrasoft contribution to $G(E)$ refers to diagrams where 
one gluon has ultrasoft momentum $k \sim E \sim m
v^2$, while an infinite number of potential (Coulomb) 
gluons can be exchanged between the heavy quarks, promoting 
the heavy-quark propagators to the Green function of the 
Schr\"odinger operator including the colour Coulomb potential. Computing 
Feynman integrals with Coulomb Green functions while simultaneously 
regulating all divergences dimensionally to be consistent with 
fixed-order matching calculations (defined according to the 
threshold expansion \cite{BenSmi}) is the main challenge of the ultrasoft 
calculation. From the ultrasoft correction to (\ref{defG}), 
$\delta^{us} G(E)$ (Eqns. (8), (9) of \cite{Beneke:2007pj}), the 
corresponding correction to the inclusive heavy-quark production 
cross section is computed as 
\begin{equation}
[R\,]_{us} = 12\pi e_Q^2 \cdot \frac{3}{2 m^2}\cdot\mbox{Im}\;
\delta^{us} G(E),
\label{R}
\end{equation}
where $R=\sigma_{Q\bar QX}/\sigma_0$ with $\sigma_0=4\pi\alpha_{\rm
  em}^2/(3 s)$ is the usual $R$-ratio, and $e_Q$ the heavy-quark
electric charge in units of the positron charge.

The calculation of the ultrasoft contribution to the 
current-correlation function involves ultraviolet divergences of 
various kinds. Divergences in box-type subdiagrams that do not contain 
any of the two current operator insertions are subtracted by 
counterterms related to the renormalization of the potentials 
in the effective Lagrangian. Another class of divergences arises 
from vertex subdiagrams with up to three loops and 
lines connecting to one of the current insertions; they are 
cancelled by the counterterms belonging to the non-relativistic 
heavy-quark current operators. These divergences have the same
structure in the  
correlation function and the bound-state residues, 
and have already been discussed in \cite{Beneke:2007pj}. 
In addition $G(E)$ has overall divergences from propagator-type 
diagrams (involving both operator insertions) with up to 
five loops, which are not relevant to the bound-state pole parameters. 
After subtraction of subdivergences the overall divergences are 
polynomial in the external ``momentum'' $E$ and should not 
contribute to the imaginary part (\ref{R}). However, there is 
a subtlety for top-quark pair production, which we now explain 
(see a related discussion in \cite{Hoang:2004tg}).

\begin{figure}[t]
  \begin{center}
  \vspace*{-0.9cm}
  \includegraphics[width=1\textwidth]{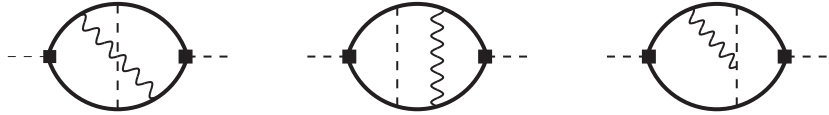}
  \vspace*{-19.7cm}
  \caption{Three-loop diagrams generating an overall divergence 
  proportional to  $\alpha_s^2 E/\epsilon$. Thick lines denote 
  an unstable top-quark propagator, wavy (dashed) lines ultrasoft 
  (potential) gluons; the square the non-relativistic current 
  insertion. Symmetric versions of the last two diagrams are not 
  displayed.}
  \label{fig1}
  \end{center}
\end{figure}

\begin{figure}[b]
  \begin{center}
  \vspace*{-0.9cm}
  \includegraphics[width=1\textwidth]{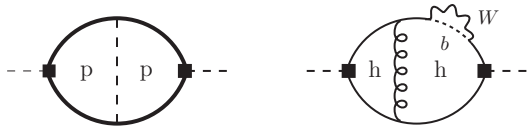}
  \vspace*{-19.7cm}
  \caption{Cancellation of the overall divergence $\alpha_s
    E/\epsilon$ at NNLO against a partially ``non-resonant'' electroweak 
    contribution involving a $Wb$ top-quark self energy insertion. 
    Symmetric versions of the last diagram are not 
   displayed.}
  \label{fig2}
  \end{center}
\end{figure}

The issue for top quarks is its large decay width $\Gamma_t \sim 
1.4\,\mbox{GeV}$, which happens to be of order of the relevant 
non-relativistic energies $E$. The correct prescription for the 
calculation of the ultrasoft correction is to replace $E\to 
E+i\hspace{0.03cm}\Gamma_t$ in all equations \cite{FK87}. Equivalently, we 
may consider $E$ to be complex with a finite imaginary part 
rather than the $+i\epsilon$-prescription for stable quarks. Now, 
the ultrasoft calculation yields an overall divergence (arising 
from quadratically divergent three-momentum integrals) 
\begin{equation}
[\delta^{us} G(E)]_{\rm overall} \propto 
\frac{\alpha_s^2}{\epsilon}\cdot E
\end{equation}
from the three-loop diagrams shown in Figure~\ref{fig1}. 
 If $E$ is complex, 
this results in a divergence $\alpha_s^2\Gamma_t/\epsilon$ 
in the heavy-quark production cross section (\ref{R}) that is 
not cancelled by any counterterm associated with the 
heavy-quark currents. A similar overall 
divergence has already appeared in NNLO calculations of 
top-quark pair production such as \cite{Beneke:1999qg}, where 
it arises from two-loop diagrams with both loops 
in the potential region as shown in the first diagram in 
Figure~\ref{fig2}. The origin and cancellation of such divergences 
is best understood in unstable-particle effective 
field theory \cite{Beneke:2003xh}, which provides a consistent 
treatment of finite-width effects beyond the $E\to 
E+i\hspace{0.03cm}\Gamma_t$ replacement (valid only up to NLO),  
and includes non-resonant 
contributions to physical cross sections, here $e^+e^-\to W^+
W^- b\bar {b}$, which do not include the unstable top 
quark in the final state. In this framework, the 
overall ultraviolet divergence $\alpha_s \Gamma_t/\epsilon$ from the first 
diagram in Figure~2 is cancelled against an infrared divergence 
from the second diagram. Note that the second diagram is not present 
in non-relativistic QCD or for stable quarks, since all loops 
are in the hard region, and that the electroweak self-energy insertion 
is computed in conventional perturbation theory in the full 
standard model. The second diagram is ${\cal O}(\alpha_s \alpha_w)$, 
so the parameter dependence matches, since 
$\Gamma_t \propto \alpha_w$, where $\alpha_w$ denotes the 
SU(2) gauge coupling. A similar cancellation between the overall 
divergences from Figure~\ref{fig1} with non-resonant 
contributions is expected at NNNLO. 

The ultrasoft contribution to the NRQCD correlation function 
provides only part of the third-order result for the heavy-quark pair
production cross section near threshold, and is 
regularization- and scheme-dependent. Our conventions for 
dimensional regularization and the counterterms have been 
specified in \cite{Beneke:2007pj} and are chosen to conform 
with standard conventions for the calculation of 
$\overline{\rm MS}$-subtracted coefficient functions. 
With respect to the overall
divergences discussed above, we note that we perform the 
calculation of the correlation function 
(\ref{defG}) with NRQCD for stable quarks in the complex 
energy plane at $E+i\hspace{0.03cm}\Gamma_t$. This corresponds to keeping 
only the leading width-correction in the effective Lagrangian 
for unstable quarks, $(i\Gamma_t/2) \,\psi^\dagger\psi$, 
in the framework  \cite{Beneke:2003xh}. The $1/\epsilon$ poles 
are then subtracted minimally ($\overline{\rm MS}$). The resulting 
scheme and scale dependence cancels with other NRQCD contributions 
to the correlation function, and matching coefficients, but 
there is a left-over dependence proportional to $\Gamma_t$ due to 
the overall divergences discussed above, which 
cancels with electroweak corrections that are not yet known.

We are now in the position to present our main result. (The 
technical details of the calculation will be explained together 
with those of \cite{Beneke:2007pj} in a separate publication.) 
We express this in terms of the dimensionless complex  
energy variable $\hat E\equiv E/(m \alpha_s^2)$, where 
$\mbox{Im}\,\hat E = \hat\Gamma_t=\Gamma_t/(m \alpha_s^2)$, and 
\begin{equation}
\lambda=\frac{C_F}{2 \sqrt{-\hat E}},
\qquad 
{\cal P} = \ln\left(\frac{C_F}{\lambda}\right)
+\gamma_{E} +\psi(1-\lambda), 
\end{equation}
with $C_F=4/3$, and $\psi(z)\equiv d\Gamma(z)/dz$ the $\psi$-function 
($\psi^{\,\prime}(z)$ denotes its first
derivative). $\hat G_C=(2/3)(1-1/\lambda-2\,{\cal P})$, used below,  
is related to the Coulomb Green function at zero radial distance by 
$G_C = m^2\alpha_s/(4\pi)\,\hat G_C$. 
The imaginary part of the correlation function at complex energy 
involves divergent, logarithmic and finite contributions. We 
obtained the divergent and logarithmic terms in closed form 
and the remainder numerically.\footnote{Dropping the divergent 
$1/\epsilon$ pole terms in (\ref{result}) 
amounts to $\overline{\rm MS}$ renormalization.} 
We thus obtain
\begin{eqnarray}
\delta^{us} G(E) &=& 
\frac{2 m^2 \alpha_s^4}{9 \pi^2} 
\Bigg\{
\,
\bigg[
\frac{17\,i \hat \Gamma_t}{24}
+\frac{527\,\hat G_C}{72}
\bigg]\,\frac{1}{\epsilon^2}
+ \,
\bigg[
\frac{17\,i \hat \Gamma_t}{12}
+\frac{221\,\hat G_C}{36}
\bigg]
\,\frac{1}{\epsilon}\,\ln\frac{\mu}{m} 
\nonumber\\
&& \hspace*{-1.5cm} 
+ \,
\bigg[\bigg(
\frac{19}{12}\,\ln 2
-\frac{91}{72}\bigg)\,i \hat \Gamma_t 
+\bigg(-\frac{119}{12}\,\ln 2
+\frac{2059}{108}\bigg)\,
\hat G_C\bigg] \,\frac{1}{\epsilon}
\nonumber\\
&& \hspace*{-1.5cm} 
+ \,
\bigg[
-\frac{34\,i \hat \Gamma_t}{3}
-\frac{595\,\hat G_C}{9}
\bigg]\,\ln^2\alpha_s
+ 
\bigg[
-\frac{17\,i \hat \Gamma_t}{12}
-\frac{833\,\hat G_C}{36}
\bigg]
\,\ln^2\frac{\mu}{m}
\nonumber \\
&& \hspace*{-1.5cm}
+ \,
\bigg[
\frac{34\,i \hat \Gamma_t}{3}
+\frac{748\,\hat G_C}{9}
\bigg]
\,\ln\alpha_s\ln\frac{\mu}{m}
\nonumber\\
&& \hspace*{-1.5cm} 
+ \,
\bigg[
\frac{2380\,{\cal P}^2}{27} 
+\left(\frac{272 \ln{2}}{9}
      -\frac{23483}{162}
      +\frac{2380}{27 \lambda}
      +\frac{272}{27 \lambda^2}
\right) {\cal P}
+\left(\frac{27 \lambda}{2}-\frac{16}{3 \lambda }
 \right) \psi^{\,\prime}(1-\lambda)
\nonumber \\
&& \hspace*{-1cm}
+
\frac{64}{27 \lambda ^3}
+\frac{4\, (-1331+306 \ln{2})}{81 \lambda}
+\frac{4\, (-199+114 \ln{2})}{81 \lambda^2}
\bigg] \,\ln\alpha_s
\nonumber \\
&& \hspace*{-1.5cm}
+ 
\,\bigg[
-\frac{1496\,{\cal P}^2}{27} 
+\left(-\frac{34 \ln{2}}{3}
       +\frac{5065}{72}
       -\frac{1496}{27 \lambda }
       -\frac{136}{27 \lambda^2}
\right) {\cal P}
+
\left(\frac{8}{3 \lambda }
-\frac{81 \lambda }{8}\right) \psi^{\,\prime}(1-\lambda)
\nonumber \\
&& \hspace*{-1cm}
-\frac{32}{27 \lambda^3}
+ \frac{163-114 \ln{2}}{27\lambda^2}
+\frac{271-51 \ln{2}}{9 \lambda }
\bigg]
\,\ln\frac{\mu}{m}
+ \, 
\delta^{us}(\hat E)\Bigg\}.
\label{result}
\end{eqnarray}
The double logarithmic and $1/\epsilon$ pole
terms are identical to those that appear in the result for the 
wave function at the origin (Eqn.~(13) of \cite{Beneke:2007pj}) under the 
replacements $\Gamma_t\to 0$, $\hat G_C \to 
K\equiv (8/9)\,(\alpha_s^3/\pi) \,|\psi_n^{C}(0)|^2/\left(2 m^2 \alpha_s^4/(9
  \pi^2)\right)$. By expanding $\delta^{us} G(E)$ around the bound-state 
poles $\lambda=n$, we also reproduce the single logarithmic terms 
in \cite{Beneke:2007pj}.\footnote{To this end write
\begin{equation}
\delta^{us} G(E) \,\stackrel{\lambda\to n}{=} \,
\frac{a_n}{(n-\lambda)^2}+\frac{b_n}{n-\lambda}+\ldots.
\end{equation}
The correction to $|\psi_n(0)|^2$ is then given by 
$(3K/4)\,(b_n+3 a_n/n)$.}
Only the imaginary part of $\delta^{us} G(E)$ 
is needed for the cross section (\ref{R}), therefore in writing 
(\ref{result}) we already dropped some purely real terms. 
The imaginary part of the non-logarithmic correction 
$\delta^{us}(\hat E)$ is tabulated in Table~\ref{tab:numbers} 
for a set of values of the real part (rows) and imaginary part 
(columns) of $\hat E$ in a range relevant to top-quark pair production.  
The table can be used to generate an interpolating function 
with an accuracy better than 0.1 in the entire range of 
the table.\footnote{For instance, using {\tt Mathematica}'s built-in function
{\tt Interpolation} with default setting.}

\begin{table}[p]
\begin{tabular}{cc}
\hspace*{-1.2cm}
{\small 
\begin{tabular}{c|ccccc}
$\mbox{Re} \,\hat E\;\backslash \;\hat \Gamma_t$  
& 0.343& 0.443& 0.543& 0.643& 0.743 \\
\hline 
-4.10  & 2.282  & 2.790  & 3.182  & 3.432  & 3.521 \\ 
-3.90  & 5.542  & 6.966  & 8.247  & 9.357  & 10.27 \\ 
-3.70  & 9.255  & 11.72  & 14.00  & 16.07  & 17.91 \\ 
-3.50  & 13.53  & 17.18  & 20.60  & 23.76  & 26.62 \\ 
-3.30  & 18.51  & 23.53  & 28.26  & 32.66  & 36.68 \\ 
-3.10  & 24.40  & 31.01  & 37.26  & 43.08  & 48.42 \\ 
-2.90  & 31.46  & 39.97  & 47.99  & 55.45  & 62.28 \\ 
-2.70  & 40.09  & 50.87  & 61.00  & 70.37  & 78.90 \\ 
-2.50  & 50.87  & 64.42  & 77.06  & 88.66  & 99.13 \\ 
-2.30  & 64.68  & 81.65  & 97.33  & 111.5  & 124.2 \\ 
-2.10  & 82.90  & 104.2  & 123.6  & 140.8  & 155.8 \\ 
-1.90  & 107.8  & 134.7  & 158.5  & 179.1  & 196.5 \\ 
-1.70  & 143.6  & 177.5  & 206.5  & 230.5  & 249.8 \\ 
-1.50  & 197.8  & 240.5  & 274.9  & 301.4  & 320.8 \\ 
-1.30  & 285.8  & 338.2  & 375.8  & 400.8  & 415.8 \\ 
-1.10  & 441.4  & 497.2  & 527.3  & 539.5  & 540.1 \\ 
-1.00  & 565.1  & 612.6  & 628.2  & 625.5  & 612.9 \\ 
-0.96  & 627.4  & 667.2  & 673.7  & 662.6  & 643.4 \\ 
-0.92  & 698.6  & 727.1  & 721.8  & 701.1  & 674.4 \\ 
-0.88  & 779.8  & 792.1  & 772.3  & 740.4  & 705.6 \\ 
-0.84  & 871.7  & 862.0  & 824.6  & 780.1  & 736.6 \\ 
-0.80  & 974.9  & 936.0  & 877.8  & 819.6  & 767.0 \\ 
-0.76  & 1089.  & 1013.  & 930.7  & 858.1  & 796.4 \\ 
-0.72  & 1211.  & 1089.  & 982.1  & 894.8  & 824.3 \\ 
-0.68  & 1336.  & 1164.  & 1030.  & 928.8  & 850.1 \\ 
-0.64  & 1458.  & 1231.  & 1073.  & 959.3  & 873.5 \\ 
-0.60  & 1565.  & 1288.  & 1109.  & 985.3  & 893.9 \\ 
-0.56  & 1643.  & 1329.  & 1137.  & 1006.  & 911.2 \\ 
\end{tabular}
}
&
{\small 
\begin{tabular}{c|ccccc}
$\mbox{Re} \,\hat E\;\backslash \;\hat \Gamma_t$ 
& 0.343& 0.443& 0.543& 0.643& 0.743 \\
\hline 
-0.52  & 1681.  & 1352.  & 1155.  & 1022.  & 925.0 \\ 
-0.48  & 1673.  & 1356.  & 1163.  & 1032.  & 935.4 \\ 
-0.44  & 1620.  & 1341.  & 1162.  & 1037.  & 942.4 \\ 
-0.40  & 1534.  & 1311.  & 1153.  & 1036.  & 946.2 \\ 
-0.36  & 1433.  & 1271.  & 1138.  & 1032.  & 947.1 \\ 
-0.32  & 1332.  & 1226.  & 1118.  & 1024.  & 945.7 \\ 
-0.28  & 1243.  & 1180.  & 1095.  & 1014.  & 942.1 \\ 
-0.24  & 1169.  & 1137.  & 1072.  & 1002.  & 937.1 \\ 
-0.20  & 1112.  & 1099.  & 1049.  & 989.3  & 931.0 \\ 
-0.16  & 1070.  & 1065.  & 1027.  & 976.4  & 924.0 \\ 
-0.12  & 1038.  & 1037.  & 1007.  & 963.7  & 916.7 \\ 
-0.08  & 1015.  & 1014.  & 989.0  & 951.4  & 909.2 \\ 
-0.04  & 996.8  & 994.4  & 972.9  & 939.9  & 901.7 \\ 
0.00  & 982.4  & 978.0  & 958.7  & 929.2  & 894.4 \\ 
0.10  & 955.4  & 946.6  & 929.7  & 905.8  & 877.4 \\ 
0.30  & 918.9  & 906.4  & 890.6  & 871.7  & 850.1 \\ 
0.50  & 893.4  & 880.3  & 865.3  & 848.7  & 830.4 \\ 
0.70  & 874.4  & 861.5  & 847.4  & 832.2  & 816.0 \\ 
0.90  & 859.8  & 847.3  & 834.0  & 819.9  & 805.2 \\ 
1.10  & 848.3  & 836.4  & 823.8  & 810.7  & 797.0 \\ 
1.30  & 839.2  & 827.8  & 816.0  & 803.6  & 790.9 \\ 
1.50  & 832.2  & 821.3  & 810.0  & 798.3  & 786.3 \\ 
1.70  & 826.7  & 816.3  & 805.5  & 794.4  & 783.1 \\ 
1.90  & 822.5  & 812.5  & 802.3  & 791.7  & 781.0 \\ 
2.10  & 819.5  & 810.0  & 800.1  & 790.1  & 779.8 \\ 
2.30  & 817.6  & 808.3  & 798.9  & 789.3  & 779.4 \\ 
2.50  & 816.5  & 807.6  & 798.5  & 789.3  & 779.8 \\ 
2.70  & 816.2  & 807.6  & 798.9  & 790.0  & 780.9 \\ 
\end{tabular}
}
\end{tabular}
\vskip0.2cm
\caption{\label{tab:numbers}
Value of the non-logarithmic ultrasoft correction 
$\mbox{Im}\,\delta^{us}(\hat E)$ 
for various scaled energies $\hat E=E/(m \alpha_s^2)$. Columns refer 
to different $\mbox{Re}\,\hat E$ from $-4.1$ to $2.7$, rows 
to five values of $\mbox{Im} \,\hat E = \hat\Gamma_t=
\Gamma_t/(m \alpha_s^2)$ for given real part.}
\end{table}

\begin{figure}[ht]
  \begin{center}
  \vspace*{1cm}
  \includegraphics[width=0.55\textwidth]{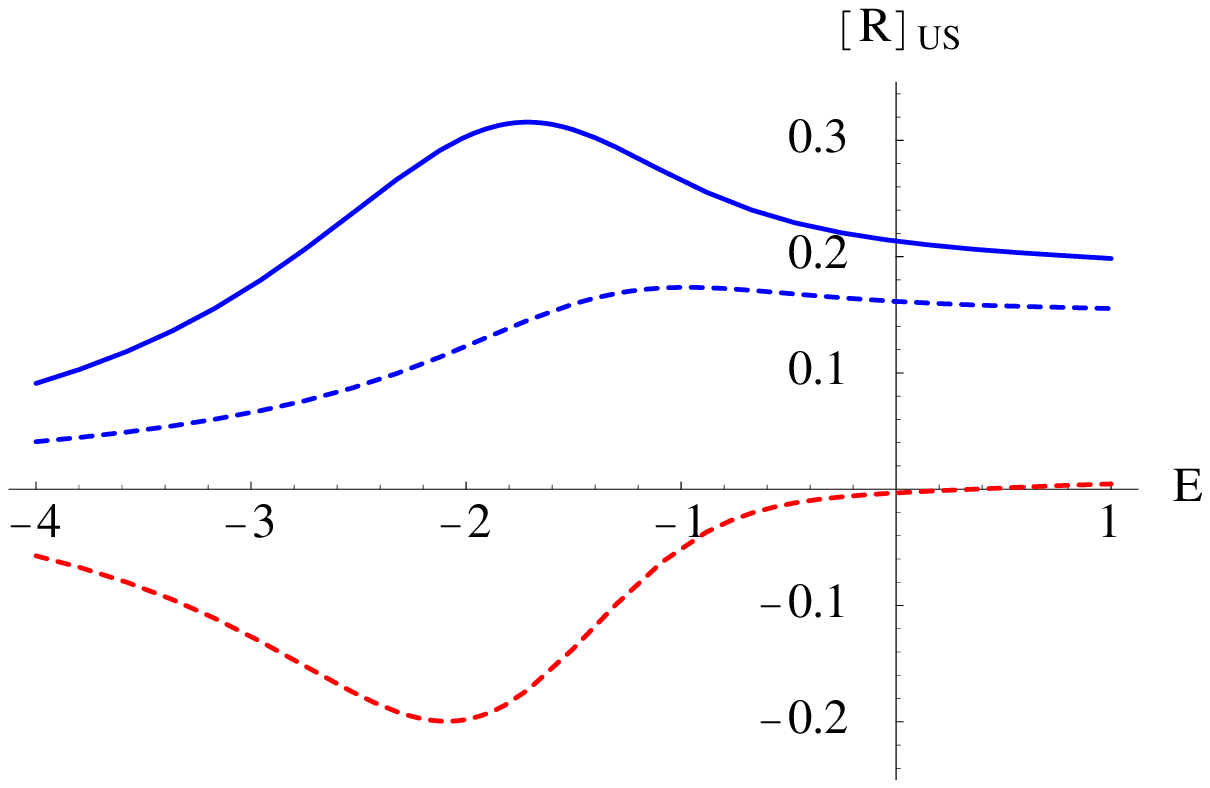}
  \vskip1cm
  \includegraphics[width=0.55\textwidth]{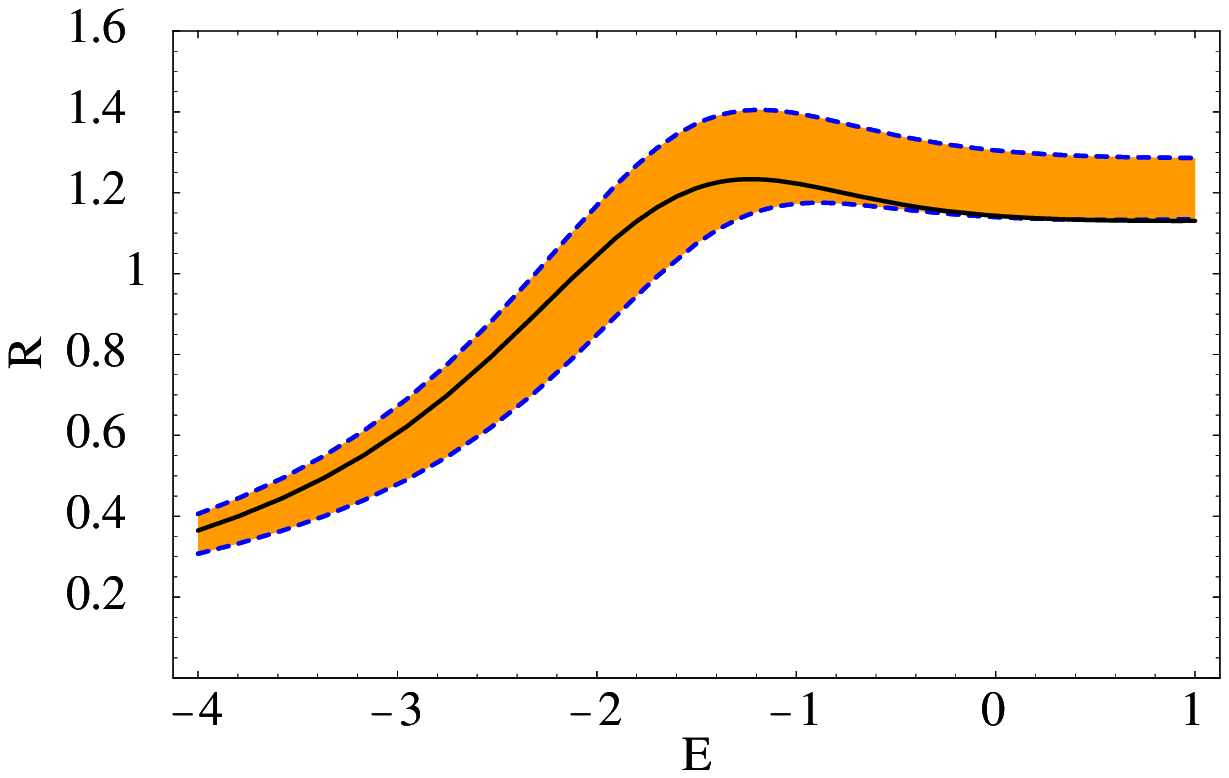}
  \caption{Top panel: Ultrasoft correction $[R]_{us}$ 
  to $t\bar t$ production as 
  function of $E=\sqrt{s}-2 m_t$. Solid: non-logarithmic contribution 
  only. Dashed: total contribution with $\mu=\mu_B=32.6\,$GeV (upper 
  dashed) and $\mu=m_t=175\,$GeV (lower dashed). Parameters: 
  $m_t=175\,$GeV, $\Gamma_t=1.4\,$GeV, $\alpha_s=0.14$. 
  Bottom panel: Ultrasoft correction added to the leading order (LO)
  cross section, i.e. $[R]_{\rm LO}+[R]_{us}$. 
  The band is obtained by varying 
  the scale $\mu$ between $\mu_B$ (upper line) to $m_t$ (lower
  line). The solid black line refers to the LO cross section.}
  \label{fig3}
  \end{center}
\end{figure}

The new ultrasoft correction has a very large effect on 
the $t\bar t$ cross section near threshold as illustrated in 
Figure~\ref{fig3}. Here we adopt the top quark 
pole mass $m_t=175\,$GeV, and fix $\alpha_s=0.14$, which corresponds 
to the Bohr radius scale $\mu_B=32.5\,$GeV. The solid line in 
the upper panel of 
Figure~\ref{fig3} shows the non-logarithmic contribution from 
$\mbox{Im}\,\delta^{us}(\hat E)$ to $[R]_{us}$ alone, which is 
seen to be around $+25\%$ in the peak region, in nice agreement 
with the estimate from the wave-function 
calculation \cite{Beneke:2007pj}. Including the logarithmic 
term requires a choice of the scale $\mu$. Since $\mu$ is not 
related to the scale of $\alpha_s$, but designates a factorization 
scale that separates the ultrasoft from hard and potential 
contributions, we choose the two values $\mu_B$ and $m_t$ 
to represent a reasonable range. Since the factorization scale 
dependence is sizable this results in a large range of 
$[R]_{us}$ reflected in the two dashed curves in 
Figure~\ref{fig3} (upper panel). We add these two results to the leading order 
$t\bar t$ cross section in the lower panel. Our results show 
that despite the large quark mass, third-order perturbative
corrections from the ultrasoft scale can have a significantly larger
impact on top-quark pair production than anticipated. However, it 
should be kept in mind that the ultrasoft correction is not 
physical by itself as is clear from its factorization scale 
dependence. (The non-logarithmic term is still factorization-scheme 
dependent.) In \cite{Beneke:2008ec} we combined the ultrasoft 
correction reported here with the third-order potential 
correction \cite{BKSunpublished} and all other calculated third-order 
terms and observed a significant cancellation between the 
ultrasoft and potential terms. A final assessment of theoretical 
uncertainties should therefore be attempted only once all third-order 
corrections to the $t\bar t$ cross section have been assembled. 

The heavy-quark correlation function near threshold is also 
a crucial input to the determination of $m_b$ from 
large-moment bottomonium sum rules \cite{VZ87}. The $n$th moment of 
the $b\bar{b}$ production cross section is defined by
\begin{eqnarray}
{\cal M}_n/(10\,\mbox{GeV})^{2 n}
&\equiv& \frac{12\pi^2}{n!}\,
\frac{d^n}{d(q^2)^n}\,\Pi(q^2)_{\big| q^2=0} = 
\int\limits_0^\infty \!
\frac{ds}{s^{n+1}}\,R_{b\bar{b}}(s).
\label{momdef}
\end{eqnarray}
Taking large $n$, typically $n \geq 4$, enhances the sensitivity to 
$m_b$ and the threshold region, but requires a non-relativistic 
treatment to sum corrections of order $\alpha_s\sqrt{n}$ 
to all orders in perturbation theory. On the other hand  
$n$ cannot be too large, since $m_b/n$, 
the typical non-relativistic energy of 
a  $b\bar{b}$ pair contributing to the $n$th moment must be 
larger than $\Lambda_{\rm QCD}$ to justify a perturbative 
computation \cite{Beneke:1999fe}. To calculate the 
derivatives of the QCD vector current correlation function 
$\Pi(q^2)$, we evaluate the moment integral in (\ref{momdef}) 
with $R_{b\bar{b}}(s)$ expressed as a sum over 
Coulomb resonances and a continuum for $E=\sqrt{s}-2 m_b >0$, 
which should be dual to the corresponding integrated 
physical $b\bar b$ cross section.

The ultrasoft correction $[R]_{us}$ to the continuum cross section 
is obtained from (\ref{R}) and 
\begin{equation}
\mbox{Im}\; \delta^{us} G(E) 
=\lim_{\Gamma \to 0+}  \,\mbox{Im}\; \delta^{us} G(E+i\hspace{0.03cm}\Gamma) 
\qquad (E>0).
\end{equation}
While we have an analytic representation of the logarithmic
contributions, see (\ref{result}), our numerical implementation 
of the non-logarithmic term $\delta^{us}(\hat E)$ does not 
allow us to make the imaginary part of $\hat E$ arbitrarily 
small. We therefore calculate 
$\delta^{us} G(E+i\hspace{0.03cm}\Gamma)$ for several values of $\Gamma$ 
(for given $E$) and obtain the continuum value by extrapolating 
to $\Gamma=0$ through a polynomial fit. By variations 
of this procedure we estimate that the relative error in the calculation 
of the ultrasoft contribution to the continuum cross section 
(real $E>0$) is less than $1\%$. Numbers for the non-logarithmic 
term are provided in Table~\ref{tab:numberscontinuum}.

\begin{table}[t]
\begin{center}
\begin{tabular}{cccc}
\hspace*{-0.0cm}
{\small 
\begin{tabular}{c|c}
$\hat E $  & $\mbox{Im}\,[\delta^{us}]$ \\
\hline 
0     & 979.0 \\ 
0.05  & 973.5 \\ 
0.10  & 968.7 \\ 
0.20  & 958.6 \\ 
0.30  & 948.3 \\ 
0.40  & 938.3 \\ 
0.50  & 928.7 \\ 
\end{tabular}
}
&
{\small 
\begin{tabular}{c|c}
$\hat E $  & $\mbox{Im}\,[\delta^{us}]$ \\
\hline 
0.70  & 911.5 \\ 
0.90  & 896.9 \\ 
1.10  & 884.7 \\ 
1.30  & 874.7 \\ 
1.50  & 866.5 \\ 
1.70  & 859.9 \\ 
1.90  & 854.6 \\ 
\end{tabular}
}
&
{\small 
\begin{tabular}{c|c}
$\hat E $  & $\mbox{Im}\,[\delta^{us}]$ \\
\hline 
2.10  & 850.5\\ 
2.50  & 845.0 \\ 
3.10  & 843.6 \\ 
3.70  & 848.0 \\ 
4.30  & 857.8 \\ 
4.90  & 871.3 \\ 
5.50  & 888.0 \\ 
\end{tabular}
}
&
{\small 
\begin{tabular}{c|c}
$\hat E $  & $\mbox{Im}\,[\delta^{us}]$ \\
\hline 
6.10  & 908.2 \\ 
6.80  & 935.7 \\ 
8.00  & 990.0 \\ 
9.00  & 1042.5 \\ 
10.0  & 1101.1 \\ 
12.0  & 1234.8 \\ 
14.0  & 1388.3 \\ 
\end{tabular}
}
\end{tabular}
\vskip0.2cm
\caption{\label{tab:numberscontinuum}
Values of the non-logarithmic ultrasoft correction 
$\mbox{Im}\,[\delta^{us}(\hat E+i\epsilon)]$ in the continuum 
(real positive $\hat E$).}
\end{center}
\end{table}

\begin{table}
  \begin{center}
    \begin{tabular}{l|c|c|c}
 \hline
 $n$ & 6 & 10 & 14   \\
 \hline 
 ${\cal M}_n^{\rm LO}$    & 0.134 & 0.122 & 0.139 \\
 ${\cal M}_n^{\rm LO+US}$ & 0.178 & 0.190 & 0.250 \\
 \hline
    \end{tabular}
  \caption{\label{tab2} Moments of the $b\bar b$ spectral function 
  for $m_b=4.95\,\mbox{GeV}$. (The moment integral is cut off 
at $\hat E=20$.) The renormalization/factorization 
  scale is taken 
  to be $\mu_n=2 m_b/\sqrt{n}$ corresponding to $\alpha_s(\mu_n)= 
  0.228,0.250, 0.267$ for $n=6,10,14$. The last row shows the 
  sum of the leading-order moments and the ultrasoft contribution. 
  The leading order alone is given in the first row for 
  comparison~\cite{Beneke:1999fe}.} 
  \end{center}
\end{table}

The sum of the leading-order and ultrasoft contribution to the 
moments is given in Table~\ref{tab2} together with the leading-order 
one alone for $b$-quark pole mass $m_b=4.95\,\mbox{GeV}$ and 
renormalization/factorization scale  $\mu_n=2 m_b/\sqrt{n}$. 
Given the large size of the ultrasoft term in the $t\bar t$ cross 
section it may not be surprising that here we find that the 
ultrasoft correction is 30\% to 80\% of the leading-order term, 
putting the perturbative approach into doubt for the larger 
moments. (The correction from the non-logarithmic term alone is 
even larger, cf. Figure~\ref{fig3}, upper panel.) However, as mentioned 
above, the result for the ultrasoft correction alone should be 
regarded with caution, and 
there is reason for assuming that the large ultrasoft
correction is partially a consequence of $\overline{\rm MS}$ 
factorization, such that the true third-order correction is 
smaller when all other corrections are added. 

To summarize, we evaluated the correction to the 
vector-current heavy-quark correlation function from ultrasoft 
gluon exchange, which appears first at NNNLO in the non-relativistic 
expansion, and is required for accurate 
top and bottom quark mass determinations from the threshold 
pair production cross section. The correction turns out 
to be large even for top quarks, but a definite conclusion 
on the attainable theoretical precision can only be drawn 
once the full NNNLO result, including potential and hard 
corrections, is available. A discussion of the sum of all 
known NNNLO terms for top quark production can be found 
in~\cite{Beneke:2008ec}.

\vspace*{1em}

\noindent
\subsubsection*{Acknowledgement}
We thank A.A.~Penin for helpful discussions and collaboration at an 
early stage.
This work is supported by the DFG Sonder\-forschungsbereich/Transregio~9 
``Computer\-ge\-st\"utzte Theoretische Teilchenphysik''.

\end{document}